# T-E formulation-based modeling of thin HTS shell magnetization

**Leonid Prigozhin[1*] and Vladimir Sokolovsky[2]**

[1]Blaustein Inst. for Desert Research, Ben-Gurion University of the Negev, Sde Boqer Campus 84990, Israel
[2]Physics Department, Ben-Gurion University of the Negev, Beer-Sheva 84105, Israel

*Corresponding author.

**E-mail:** leonid@bgu.ac.il and sokolovv@bgu.ac.il

## Abstract

Numerical methods for modeling thin-film magnetization are primarily focused on computing the current-density distribution. The highly nonlinear current-voltage characteristic of type-II superconductors significantly complicates the accurate computation of the electric field. The T-E formulation-based mixed finite element method, previously derived for flat superconducting films, enables the simultaneous, accurate determination of both variables. Another advantage of this method is that the computational domain is limited to the film itself: no meshing of the surrounding space is required. The thin-shell approximation reduces the problem to a two-dimensional one.

This work extends the T-E formulation and numerical method to non-flat superconducting shells with a metal substrate. We validate the method with several test examples, including modeling the magnetization of a sphere. The method is then applied to a realistic model of a cylindrical magnetic dynamo pump, and the generated open-circuit voltage is computed.

## 1. Introduction

Thin high-temperature superconducting (HTS) shells, especially thin REBCO/YBCO coated conductors shaped into cylindrical, spherical, conical, or other curved geometries, are mainly used where strong screening currents on a lightweight surface are needed. Non-flat HTS shells, either alone or in a thin multilayer assembly with ferromagnetic shells, naturally guide magnetic flux and can enclose a protected or active volume. Such shells are employed for shielding sensitive instrumentation from external magnetic fields [1, 2], to prevent the stray magnetic fields of high-power-density equipment from affecting neighboring devices [2], for magnetic cloaking [3], as tapes in multifilament cables [4], as a stator of a cylindrical magnetic flux pump [5], etc.

Different formulations and finite element methods have been developed for thin-film magnetization and transport-current problems. Since the film thickness is typically three orders of magnitude smaller than its other sizes, the H-formulation-based methods [6, 7] are less convenient for thin film modeling and, typically, cannot be applied without artificially increasing the film thickness. Using the infinitely thin film approximation is justified in most situations; it was employed, e.g., in the T and T-A formulation-based simulations, see [8-10] and [11-13], respectively. In thin-shell problems, the T-potential is the scalar stream function for the vectorial sheet current density. Using the current potential significantly simplifies the problem if the direction of the sheet current density is not known *a priori*.

Determining the electric field is needed to compute AC losses or the voltages generated by magnetic flux pumps. Accurate computation of this field, however, is difficult for most numerical methods. For Bean-like critical-state models with multivalued current-voltage relations, the electric field remains unknown even if the current density is computed. This is a limiting case, but the difficulty persists in models with more realistic, yet highly nonlinear relations, typical of type-II superconductors. Thus, although the current potential $T$ can be computed with high accuracy, the accuracy of the sheet current density, its curl, is inevitably lower. When used directly to calculate the electric field, the nonlinear current-voltage relation magnifies sheet current density errors by many times.

To overcome this difficulty, a mixed finite element method for modeling magnetization in flat HTS films has been proposed in [14]. This T-E method was based on an integral formulation in terms of two variables, the current potential and electric field, and used Raviart-Thomas elements of the lowest order for electric field $E$ (rotated) and piecewise linear continuous elements for $T$. A simpler approximation, using the vector piecewise constant elements for $E$ and piecewise linear nonconforming elements for $T$, was later proposed in [15]. We note that, when replacing the nonconforming elements in this combination with piecewise linear continuous ones, spurious oscillations in the computed electric field appear.

This work extends the T-E formulation and the finite element method [15] to non-flat HTS shells. We validate the method by solving several axisymmetric problems, assuming the power law current-voltage relation with a constant critical sheet current density, and comparing the obtained solutions with solutions by a highly accurate spectral method [16], derived especially for such problems. Then, to demonstrate the efficiency of our method in a realistic example, we consider a cylindrical magnetic dynamo pump and compute the open-circuit voltage generated in its stator. This problem is not axisymmetric. The applied magnetic field is the field of a rotating permanent magnet. In this example, we assume a field-dependent sheet critical current density and account for the substrate conductivity.

## 2. T-E formulation

In the thin-shell magnetization model employed in this work, thin shells are represented by their mid-surface $S$ characterized by a nonlinear relation between the tangential electric field $\boldsymbol{e}$ and the surface (sheet) current density $\boldsymbol{j}$. We consider a shell containing a superconducting layer on a metal substrate. For the superconducting layer, we assume a power-law resistivity, $\rho_s(\boldsymbol{j}_s) = e_0 \,|\, \boldsymbol{j}_s \,|^{n-1} / j_c^n$, where the power $n$ is a constant, the critical field $e_0 = 10^{-4}$ V/m, and the sheet critical current density $j_c$ may depend on the magnetic field and coordinates. If the sheet current density in the superconducting layer, $\boldsymbol{j}_s$, becomes overcritical, the local resistivity $\rho_s$ increases. In this area, the sheet current density in a metal substrate, $\boldsymbol{j}_m$, can become significant. Denoting by $\rho_m$ the sheet resistivity of the substrate, we write the current-voltage relation for the total current density $\boldsymbol{j} = \boldsymbol{j}_s + \boldsymbol{j}_m$ as

$$\boldsymbol{j} = \left( j_c e_0^{-1/n} \,|\, \boldsymbol{e} \,|^{-1+1/n} + \rho_m^{-1} \right) \boldsymbol{e} . \tag{1}$$

We assume $S$ is an orientable surface with the unit normal $\boldsymbol{N}(s)$ at a point $s \in S$. The surface differential operators can be defined as follows. Let $f(s)$ and $\boldsymbol{u}(s)$ be a scalar function and a tangential vector field, respectively, defined on $S$. We can consider the extensions, $\hat{f}$ and $\hat{\boldsymbol{u}}$, to a sufficiently narrow neighborhood of $S$, assuming their values remain constant along the normal direction, and set ([17], par. 2.5.6)

$$\nabla_S f := \mathrm{grad}\,\hat{f}\,|_S, \quad \nabla_S \cdot \boldsymbol{u} := \mathrm{div}\hat{\boldsymbol{u}}\,|_S, \quad \nabla_N \times \boldsymbol{u} := \boldsymbol{N} \cdot \mathrm{curl}\hat{\boldsymbol{u}}\,|_S .$$

On $S$, the Faraday law yields $\nabla_N \times \boldsymbol{e} + \partial_t b_N = 0$, where $b_N$ is the normal component of the magnetic induction. Since $b_N = \nabla_N \times \boldsymbol{a}_\tau[\boldsymbol{j}] + \mu_0 h_N^e$, where $\boldsymbol{h}^e$ is the external magnetic field and $\boldsymbol{a}_\tau$ is the tangential component of the current-induced vector potential on $S$, we obtain

$$\nabla_N \times \left(\boldsymbol{e} + \partial_t \boldsymbol{a}_\tau[\boldsymbol{j}]\right) + \mu_0 \partial_t h_N^e = 0, \tag{2}$$

where

$$\boldsymbol{a}(t,s) = \frac{\mu_0}{4\pi}\int_S \frac{\boldsymbol{j}(t,s')}{|\boldsymbol{s}-\boldsymbol{s}'|}\mathrm{d}S'. \tag{3}$$

Here and below $\boldsymbol{s}$ is the radius vector of the point $s$. If the critical sheet current density $j_c$ depends on the magnetic field $\boldsymbol{h} = \boldsymbol{h}^e + \boldsymbol{h}^j$, the shell-current-induced field on $S$,

$$\boldsymbol{h}^j(t,s) = \frac{1}{4\pi}\int_S \boldsymbol{j}(t,s') \times \frac{\boldsymbol{s}-\boldsymbol{s}'}{|\boldsymbol{s}-\boldsymbol{s}'|^3}\mathrm{d}S', \tag{4}$$

needs to be simultaneously determined. To derive a weak variational form of (2) let us multiply this equation by a differentiable on $S$ test function $\psi$ satisfying $\psi|_{\partial S} = 0$ and integrate over $S$. Since

$$\int_S \psi \nabla_N \times \boldsymbol{u}\,\mathrm{d}S = \int_S \left(\boldsymbol{u} \times \nabla_S \psi\right)\cdot \boldsymbol{N}\mathrm{dS},$$

we have $\int_S \left(\{\boldsymbol{e} + \partial_t \boldsymbol{a}_\tau\} \times \nabla_S \psi\right)\cdot \boldsymbol{N}\mathrm{d}S = -\mu_0 \int_S \psi \partial_t h_N^e \mathrm{d}S$. Noting that $\boldsymbol{N} \times \boldsymbol{a}_\tau = \boldsymbol{N} \times \boldsymbol{a}$, we obtain

$$\left(\boldsymbol{a}_\tau \times \nabla_S \psi\right)\cdot \boldsymbol{N} = \left(\boldsymbol{N} \times \boldsymbol{a}_\tau\right)\cdot \nabla_S \psi = \left(\boldsymbol{N} \times \boldsymbol{a}\right)\cdot \nabla_S \psi = -\left(\boldsymbol{N} \times \nabla_S \psi\right)\cdot \boldsymbol{a}.$$

Substituting (3), we can rewrite (2) as

$$\int_S \left(\boldsymbol{e} \times \nabla_S \psi\right)\cdot \boldsymbol{N}\mathrm{d}S - \frac{\mu_0}{4\pi}\int_S\int_S \frac{\left[\boldsymbol{N}(s) \times \nabla_S \psi(s)\right]\cdot \partial_t \boldsymbol{j}(t,s')}{|\boldsymbol{s}-\boldsymbol{s}'|}\mathrm{d}S'\mathrm{d}S = -\mu_0 \int_S \psi \partial_t h_N^e \mathrm{d}S. \tag{5}$$

In the magnetization problems, the sheet current is divergence-free, $\nabla_S \cdot \boldsymbol{j} = 0$, and also $\boldsymbol{\nu} \cdot \boldsymbol{j} = 0$ on the domain boundary $\partial S$; here $\boldsymbol{\nu}$ is a vector tangential to $S$ and normal to $\partial S$. If $S$ is a simply connected surface with a boundary, there exists the scalar T-potential (stream function) $T$ such that

$$\boldsymbol{j} = \boldsymbol{N} \times \nabla_S T \tag{6}$$

and $T|_{\partial S} = 0$. If the surface is simply connected without boundary, like a sphere, we can set, e.g., $T(s_0) = 0$ at an arbitrarily chosen point $s_0 \in S$ to ensure the stream function is unique. We further extend the set of admissible surfaces to include not simply connected surfaces $S_0$, topologically equivalent to a sphere with several holes. Denoting the holes in $S_0$ as $D_0, D_1, \ldots, D_K$, we cover them all except $D_0$ with a thin film of a fictitious material characterized by a small constant critical sheet current density $j_c^* \ll j_c$ and an infinite substrate resistivity $\rho_m$ to suppress the induced currents there. The stream function $T$ can now be defined on a simply connected surface $S = S_0 \cup D_1 \ldots \cup D_K$ with $\partial S = \partial D_0$. This class of surfaces, although not general, is sufficient for many practical applications.

It is convenient to introduce the rotated electric field $\boldsymbol{q}=-\boldsymbol{N}\times\boldsymbol{e}$ (so that $\boldsymbol{e}=\boldsymbol{N}\times\boldsymbol{q}$) and use $\boldsymbol{e}\times\nabla_S\psi\cdot\boldsymbol{N}=-\boldsymbol{q}\cdot\nabla_S\psi$. In the new variables, the T-E formulation of the thin shell magnetization problem (1), (5) can be written as

$$\int_S \boldsymbol{q}\cdot\nabla_S\psi\,\mathrm{d}S+w\left(\partial_t T,\psi\right)=\mu_0\int_S \psi\partial_t h_N^e\,\mathrm{d}S \tag{7}$$

$$\nabla_S T=\left(j_c e_0^{-1/n}\,|\boldsymbol{q}|^{-1+1/n}+\rho_m^{-1}\right)\boldsymbol{q} \tag{8}$$

with $T|_{\partial S}=0,\quad T|_{t=0}=T_0$. Here $\psi$ is any differentiable on $S$ test function satisfying $\psi|_{\partial S}=0$, and $w$ is the bilinear form

$$w(\varphi,\psi)=\frac{\mu_0}{4\pi}\int_S\int_S\frac{\left[\boldsymbol{N}(s')\times\nabla_{S'}\varphi(s')\right]\cdot\left[\boldsymbol{N}(s)\times\nabla_S\psi(s)\right]}{|\boldsymbol{s}-\boldsymbol{s}'|}\mathrm{d}S'\mathrm{d}S\,. \tag{9}$$

## 3. Numerical method

Let $S^h$ be a piecewise linear approximation of the surface $S$, constructed using its triangulation and replacement of the curved surface elements by flat triangles with the same vertices belonging to $S$. Let $\mathcal{T}^h$ be the set of these triangles, $\mathcal{E}^h$ the set of their edges, and $\mathcal{E}^h_{in}$ the subset of internal edges. We use the implicit approximation of (7) in time and, at each time step, seek $\boldsymbol{q}$ in the space $\boldsymbol{Q}^h$ of the piecewise constant vector functions, equal on each triangle $\kappa\in\mathcal{T}^h$ to a tangential vector $\boldsymbol{q}_\kappa$. For $T$ we employ the nonconforming Crouzeix-Raviart elements: an approximation to $T$ is sought in the space $V_0^h$ of scalar functions, linear in each triangular element, continuous at the midpoints of the internal edges, and zero at the midpoints of the boundary edges. Denoting the midpoint of an edge $\lambda_j\in\mathcal{E}^h$ by $s_j$ we choose the basis functions $\psi_j\in V_0^h$ such that

$$\psi_j(s_l)=\begin{cases}1 & l=j,\ \lambda_j\in\mathcal{E}^h_{in},\\ 0 & \text{otherwise.}\end{cases}$$

The function $T=\sum_{\lambda_j\in\mathcal{E}^h_{in}}T_j(t)\psi_j(s)$ has constant spatial derivatives in each element.

To solve the discretized nonlinear problem (7)-(8) at the time level $i$, we employ iterations, taking the initial value from the previous time level, $\boldsymbol{q}^{i,0}=\boldsymbol{q}^{i-1}$, and replacing $|\boldsymbol{q}^i_\kappa|^{-1+1/n}\,\boldsymbol{q}^i_\kappa$ at an $m$-th iteration by

$$|\boldsymbol{q}_\kappa^{i,m-1}|^{-1+1/n}\,\boldsymbol{q}_\kappa^{i,m-1}+\left(|\boldsymbol{q}_\kappa^{i,m-1}|_\varepsilon\right)^{-1+1/n}\left(\boldsymbol{q}_\kappa^{i,m}-\boldsymbol{q}_\kappa^{i,m-1}\right),$$

where $|\boldsymbol{u}|_\varepsilon=\sqrt{|\boldsymbol{u}|^2+\varepsilon^2}$ with a small $\varepsilon>0$. Such iterations have been employed in our previous works [14, 15]; as before, we set $\varepsilon=10^{-9}e_0$. This yields

$$\sum_{\kappa\in\mathcal{T}^h}\boldsymbol{q}_\kappa^{i,m}\cdot\nabla_S\psi_j\Big|_\kappa|\kappa|+w^h\left(\frac{T^{i,m}-T^{i-1}}{\tau_i},\psi_j\right)=\mu_0\sum_{\kappa\in\mathcal{T}^h}\frac{\left(h_N^{e,i}-h_N^{e,i-1}\right)\Big|_{o_\kappa}}{\tau_i}\int_\kappa\psi_j\mathrm{d}\kappa, \tag{10}$$

$$\nabla_S T^{i,m}|_\kappa=j_c e_0^{-1/n}\left[|\boldsymbol{q}_\kappa^{i,m-1}|^{1/n-1}\,\boldsymbol{q}_\kappa^{i,m-1}+\left(|\boldsymbol{q}_\kappa^{i,m-1}|_\varepsilon\right)^{1/n-1}\left(\boldsymbol{q}_\kappa^{i,m}-\boldsymbol{q}_\kappa^{i,m-1}\right)\right]+\rho_m^{-1}\boldsymbol{q}_\kappa^{i,m}, \tag{11}$$

where $\tau_i$ is the time step, $|\kappa|$ is the area of the element $\kappa$, $o_\kappa$ is the center of $\kappa$, the operator $\nabla_S$ acts on functions defined on flat triangular elements, and

$$w^h(\varphi,\psi)=\sum_{\kappa\in\mathcal{T}^h}\sum_{\kappa'\in\mathcal{T}^h}\left[\boldsymbol{N}(\boldsymbol{s})\times\nabla_S\varphi(\boldsymbol{s})\right]\Big|_{\kappa}\cdot\left[\boldsymbol{N}(\boldsymbol{s}')\times\nabla_{S'}\psi(\boldsymbol{s}')\right]\Big|_{\kappa'}K_{\kappa,\kappa'}$$

with $K_{\kappa,\kappa'}=\frac{\mu_0}{4\pi}\int_{\kappa}\int_{\kappa'}\frac{1}{|\boldsymbol{s}-\boldsymbol{s}'|}\mathrm{d}s'\mathrm{d}s$ (see Appendix for computation of these integrals). We use (11) to express $\boldsymbol{q}_{\kappa}^{i,m}$ via $\nabla_S T^{i,m}|_{\kappa}$,

$$\boldsymbol{q}_{\kappa}^{i,m}=\frac{|\boldsymbol{q}_{\kappa}^{i,m-1}|_{\varepsilon}^{1/n-1}\boldsymbol{q}_{\kappa}^{i,m-1}-|\boldsymbol{q}_{\kappa}^{i,m-1}|^{1/n-1}\boldsymbol{q}_{\kappa}^{i,m-1}}{|\boldsymbol{q}_{\kappa}^{i,m-1}|_{\varepsilon}^{1/n-1}+e_0^{1/n}/(\rho_m j_c)}+\frac{e_0^{1/n}/j_c}{|\boldsymbol{q}_{\kappa}^{i,m-1}|_{\varepsilon}^{1/n-1}+e_0^{1/n}/(\rho_m j_c)}\nabla_S T^{i,m}|_{\kappa}, \tag{12}$$

and substitute this into (10). Finally, we obtain

$$\sum_{\kappa\in\mathcal{T}^h}D_{\kappa}^{i,m}\nabla_S T^{i,m}|_{\kappa}\cdot\nabla_S\psi_j\big|_{\kappa}+w^h\left(T^{i,m},\psi_j\right)=F_j^{i,m} \tag{13}$$

with

$$D_{\kappa}^{i,m}=\frac{e_0^{1/n}/j_c}{|\boldsymbol{q}_{\kappa}^{i,m-1}|_{\varepsilon}^{1/n-1}+e_0^{1/n}/(\rho_m j_c)}\tau_i|\kappa|,$$

$$F_j^{i,m}=a^h(T^{i-1},\psi_j)+\mu_0\sum_{\kappa\in\mathcal{T}^h}\left(h_N^{e,i}-h_N^{e,i-1}\right)\Big|_{o_\kappa}\int_{\kappa}\psi_j\mathrm{d}\kappa-$$
$$\tau_i\sum_{\kappa\in\mathcal{T}^h}\left[\frac{|\boldsymbol{q}_{\kappa}^{i,m-1}|_{\varepsilon}^{1/n-1}\boldsymbol{q}_{\kappa}^{i,m-1}-|\boldsymbol{q}_{\kappa}^{i,m-1}|^{1/n-1}\boldsymbol{q}_{\kappa}^{i,m-1}}{|\boldsymbol{q}_{\kappa}^{i,m-1}|_{\varepsilon}^{1/n-1}+e_0^{1/n}/(\rho_m j_c)}\right]\cdot\nabla_S\psi_j\big|_{\kappa}|\kappa|$$

Let $N_{ie}$ be the number of inner edges and the $N_{ie}\times N_{ie}$ matrices $A$ and $B^{i,m}$ be defined by $A_{jl}=w^h(\psi_j,\psi_l)$, $B_{jl}^{i,m}=\sum_{\kappa\in\mathcal{T}^h}D_{\kappa}^{i,m}\nabla_S\psi_l\big|_{\kappa}\cdot\nabla_S\psi_j\big|_{\kappa}$. Then, for $T^{i,m}=\sum_{\lambda_l\in\mathcal{I}_{in}^h}C_l^{i,m}\psi_l$ equations (13) can be written as $\left[B^{i,m}+A\right]\boldsymbol{C}^{i,m}=\boldsymbol{F}^{i,m}$. Here $A$ is a symmetric, positively definite, full matrix and $B^{i,m}$ is a sparse nonnegative symmetric matrix. As a result of iterations (12)-(13) we obtain a vectorial piecewise constant approximation for the electric field $\boldsymbol{e}^i=\boldsymbol{N}\times\boldsymbol{q}^i$ and a discontinuous piecewise linear approximation for the current potential $T^i$. Convergence of these iterations has been accelerated by applying the over-relaxation: the values $\boldsymbol{q}_{\kappa}^{i,m}$ computed using (12) were, on each iteration, replaced by $\alpha\boldsymbol{q}_{\kappa}^{i,m}+(1-\alpha)\boldsymbol{q}_{\kappa}^{i,m-1}$ with $\alpha=1.8$.

Although the direct computation of the piecewise constant approximation to the sheet current density as $\boldsymbol{j}^i=\boldsymbol{N}\times\nabla_S T^i$ is possible, the result may be noisy. Hence, we followed the approach proposed in [15] for the transport current problems: a better approximation was obtained as $\boldsymbol{j}^i=\boldsymbol{N}\times\nabla_S\tilde{T}^i$, where $\tilde{T}^i$ is the approximation of $T^i$ by a continuous function, linear on each element, zero at the boundary nodes, and minimizing

$$\sum_{\kappa}|\kappa|\,\|(\nabla_S\tilde{T}^i-\nabla_S T^i)|_{\kappa}\|^2. \tag{14}$$

The inner node values of $\tilde{T}^i$ that minimize (14) are obtained by solving a linear algebraic system with a sparse, positive-definite matrix.

To account for a $j_c$ dependence on $\boldsymbol{h}$, the magnetic field $\boldsymbol{h}=\boldsymbol{h}^e+\boldsymbol{h}^j$ on $S^h$ must be calculated. We used the piecewise constant approximation of the current-induced magnetic field (4), setting on each iteration (the indices of the time level and iteration number are omitted)

$$\boldsymbol{h}^j\Big|_\kappa = \sum_{\kappa'\in\mathcal{T}^h} \boldsymbol{j}_{\kappa'} \times \boldsymbol{u}_{\kappa,\kappa'} \tag{15}$$

with the analytically calculated (see Appendix) integrals

$$\boldsymbol{u}_{\kappa,\kappa'} = \frac{1}{4\pi}\int_{\kappa'} \frac{\boldsymbol{o}_\kappa - \boldsymbol{s}}{|\boldsymbol{o}_\kappa - \boldsymbol{s}|^3}\mathrm{d}s. \tag{16}$$

Our simulations were performed in MATLAB R2020b on a PC with the Intel i7-10700 CPU and 64 GB RAM. The tolerances in iterations (10)-(11) were $10^{-4}$ for $T$ and $5\cdot 10^{-4}$ for $\boldsymbol{q}$ (in the $L^1$-norm).

## 4. Validation

To validate our method, we solved three axisymmetric problems and compared their solutions with those obtained by the highly accurate Chebyshev spectral method [16], developed especially for problems where the surface $S$ is generated by the rotation of a curve around the $z$ -axis and $\boldsymbol{j} = j(t,r,z)\boldsymbol{i}_\varphi$, $\boldsymbol{e} = e(t,r,z)\boldsymbol{i}_\varphi$ in cylindrical coordinates $(r,\varphi,z)$. In this section, we present the results using scaled dimensionless variables

$$\tilde{\boldsymbol{j}} = \frac{\boldsymbol{j}}{j_c}, \quad \tilde{\boldsymbol{h}} = \frac{\boldsymbol{h}}{j_c}, \quad \tilde{\boldsymbol{e}} = \frac{\boldsymbol{e}}{e_0}, \quad \tilde{\boldsymbol{s}} = \frac{\boldsymbol{s}}{l}, \quad \tilde{t} = \frac{t}{t_0},$$

where $j_c$ is the field-independent value of the critical sheet current density, $l$ is the characteristic shell size, and $t_0 = l\mu_0 j_c / e_0$. Below, the sign tilde will be omitted for simplicity.

We assume $\boldsymbol{j}(0,s) = \boldsymbol{0}$ and a growing uniform external magnetic field is parallel to the $z$ -axis, with $h_z^e = 6t$ (the rate of magnetic field growth affects mainly the electric field magnitude). We set $n = 30$ and $\rho_m = \infty$ in the current-voltage relation (8).

The three surfaces, a hemisphere, a sphere, and a cylinder of a finite height, were chosen as examples of simply connected surfaces with a boundary, closed simply connected surfaces, and multiply connected surfaces, respectively.

The spectral method [16] uses a parametric representation of the axisymmetric surface generator, $\varsigma = \{r = R(p), z = Z(p)\}$ for $-1 \le p \le 1$, and computes the solution at $N^{Ch}$ Chebyshev points of the first kind, $p_k = \cos\left(\pi(k - 1/2)/N^{\mathrm{Ch}}\right)$, $k = 1,2,\ldots,N^{\mathrm{Ch}}$. We use the interpolating expansion in Chebyshev polynomials to calculate the approximate solution at any $p \in (-1,1)$. Finding the spectral solution values of $\boldsymbol{j}$ and $\boldsymbol{e}$ at the element centers $o_\kappa$, we compare them with the values $\boldsymbol{j}\,|_\kappa$ and $\boldsymbol{e}\,|_\kappa$, respectively, of the finite element method solution and evaluate the accuracy of the mixed method.

For the hemisphere of radius 1, computing the solution at $h_z^e = 0.6$ by the spectral method with 400 Chebyshev points took 5 seconds; the maximal absolute errors were, approximately, $2\cdot 10^{-6}$ for $\boldsymbol{j}$ and $1.5\cdot 10^{-5}$ for $\boldsymbol{e}$. The finite element solution was computed for two-dimensional meshes with 1299 elements (Fig. 1) and 5154 elements. The obtained solutions are shown in Fig. 2.

Using the spectral solution as a reference, we estimated the relative errors $\delta\boldsymbol{j}$ and $\delta\boldsymbol{e}$ of our finite element method in the $L^2$-norm (Table 1). Our mixed method uses (10)-(11) to simultaneously determine both variables, $T$ and $\boldsymbol{e} = \boldsymbol{N}\times\boldsymbol{q}$. In other methods, such as in [11, 12, 18, 19], the potential $T$ is found first, then used to calculate the sheet current density, and finally, the dimensionless electric field can be found as $\boldsymbol{e} = |\boldsymbol{j}|^{n-1}\boldsymbol{j}$. Instead of the simultaneous determination of both variables, we can follow the approach [11, 12, 18, 19] and substitute the calculated $\boldsymbol{j}$ into the current-voltage relation directly for the electric field calculation; the error of such an alternative calculation is denoted as $\delta\boldsymbol{e}^*$.

**Table 1.** Hemisphere. Relative errors and computation time of the finite element method.

| Time step | Number of elements | $\delta\boldsymbol{j}$, % | $\delta\boldsymbol{e}$, % | $\delta\boldsymbol{e}^*$, % | CPU time, min. |
|---|---|---|---|---|---|
| 1e-3 | 1299 | 1.7 | 2.2 | 5.6 | 1.9 |
| 0.5e-3 | 5154 | 0.8 | 1.2 | 2.9 | 57 |

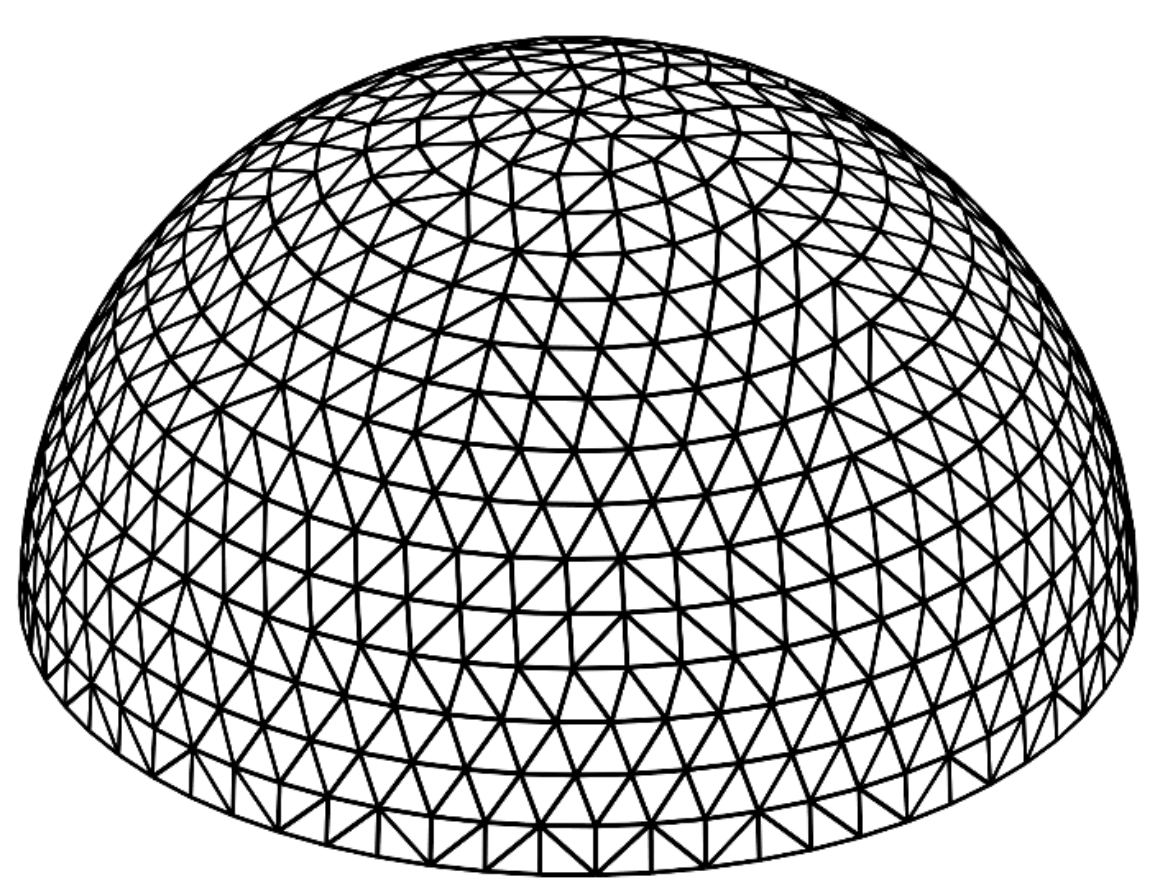

**Figure 1.** The 1299-element mesh for the hemisphere.

**Figure 2.** Simulation results for the hemispherical shell. Spectral solution (red line) and the finite element solution (black dots) for the mesh with 5154 elements. $n = 30$, $h_z^e = 0.6$, $\theta$ is the angle between the z-axis and the direction from the sphere center to a surface point (the polar angle in spherical coordinates).

The spherical shell in our next example has no boundary, and to ensure the uniqueness of the stream function, we set $T = 0$ at the midpoint of one mesh edge. This problem was solved using two meshes, with 1902 and 4046 elements (Fig. 3); the solutions at $h_z^e = 0.9$ were compared with the highly accurate solution by the spectral method; see Table 2 and Fig. 4.

**Table 2.** Sphere. Relative errors and computation time of the finite element method.

| Time step | Number of elements | $\delta\boldsymbol{j}$, % | $\delta\boldsymbol{e}$, % | $\delta\boldsymbol{e}^*$, % | CPU time, min. |
|---|---|---|---|---|---|
| 1.5e-3 | 1902 | 1.9 | 2.8 | 14 | 3.7 |
| 7.5e-4 | 4046 | 1.3 | 1.4 | 9.2 | 30 |

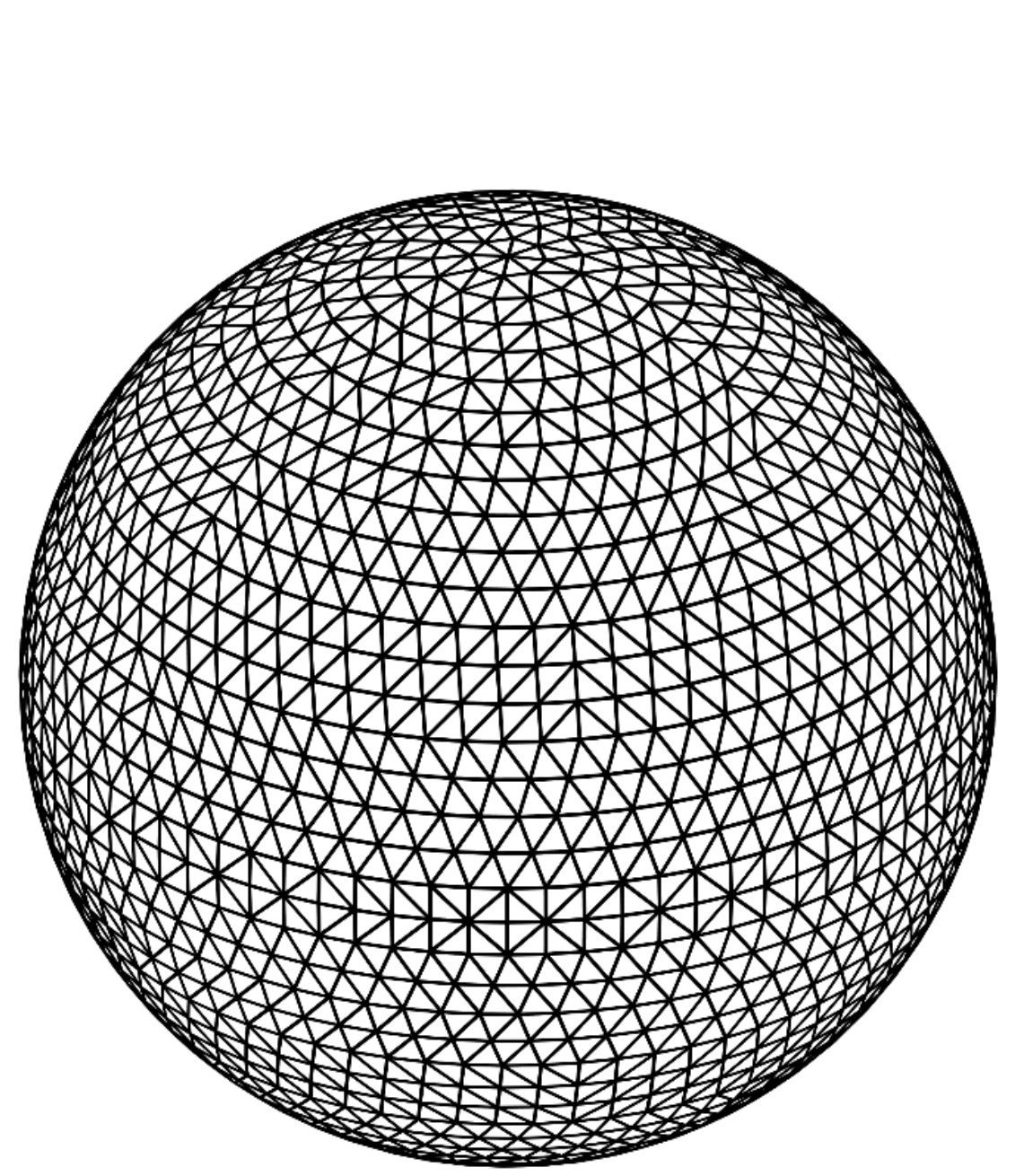

**Figure 3.** The 4046-element mesh for the spherical shell.

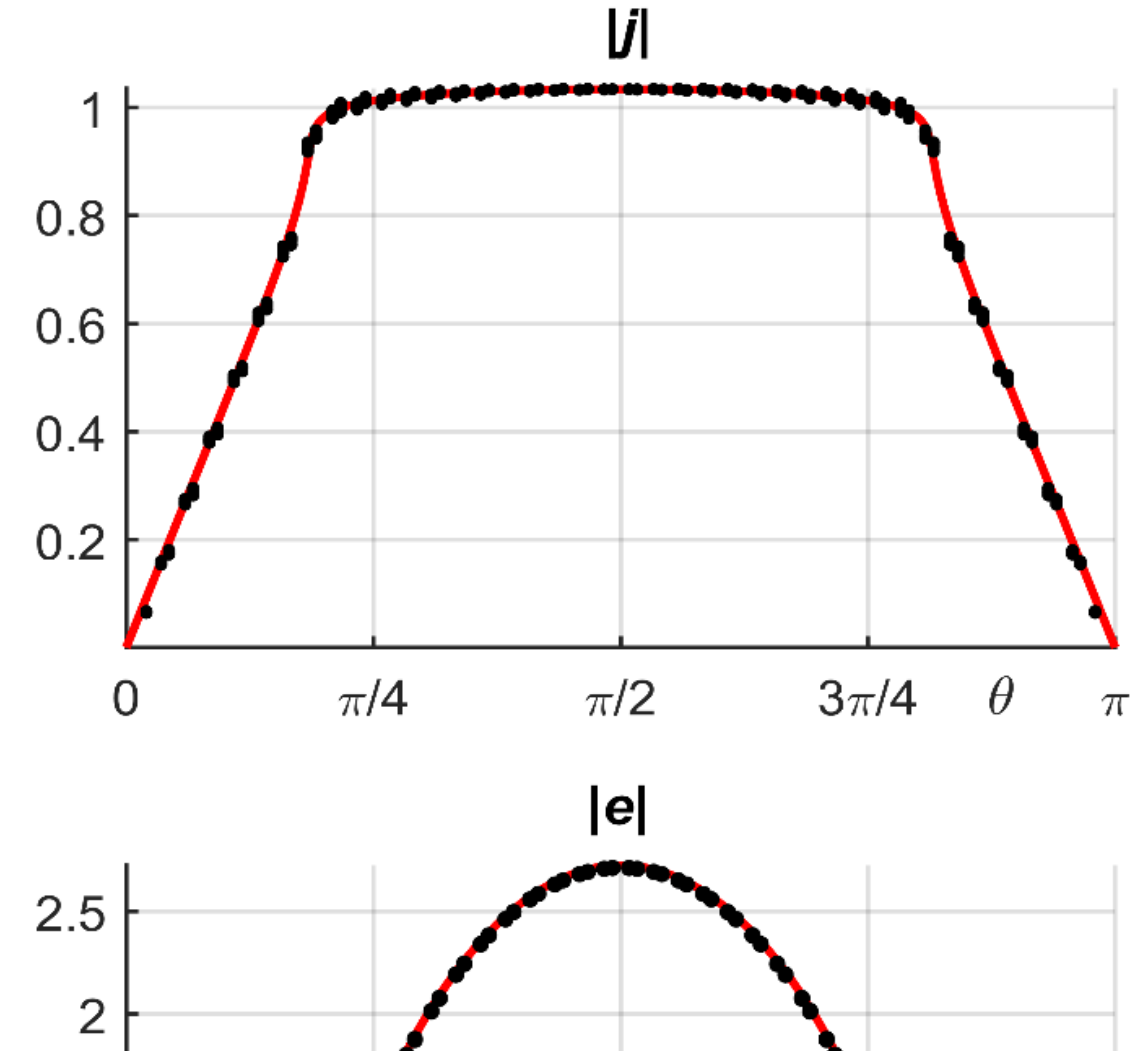

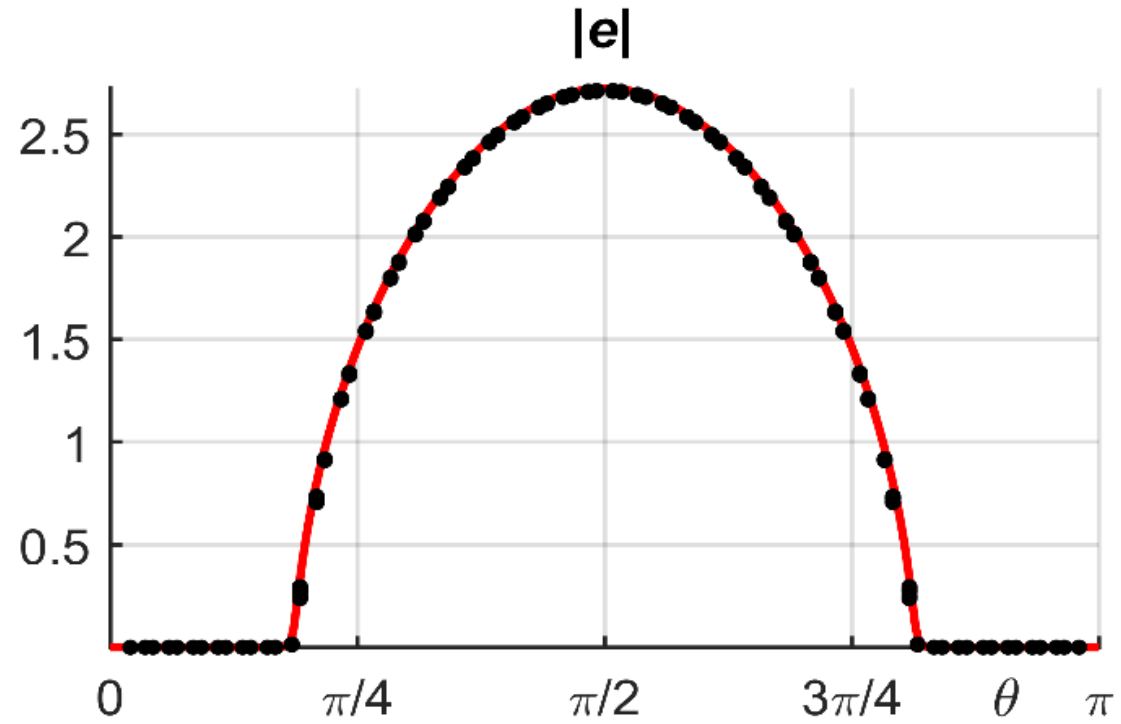

**Figure 4.** Simulation results for the spherical surface. Spectral solution (red line) and the finite element solution (black dots) for the mesh with 4046 elements. $n=30,\ h_z^e=0.9,\ \theta$ is the polar angle.

The cylindrical surface $S_0=\{r=1, |z|\le 1, 0\le\varphi<2\pi\}$ is not simply connected; topologically, it is equivalent to a sphere with two holes. To use the stream function, we close $S_0$ from one side by the circle $D_1=\{|r|<1, z=-1, 0\le\varphi<2\pi\}$ and triangulate the surface $S=S_0\cup D_1$ (Fig. 5). We set $j_c=10^{-3}$ in $D_1$ to suppress currents in this domain, keeping the dimensionless $j_c=1$ in $S_0$. The accuracy of $\boldsymbol{j}$ and $\boldsymbol{e}$ in $S_0$ at $h_z^e=0.6$ was estimated by comparing them with the spectral solution (see Table 3 and Fig. 6).

**Table 3.** Cylinder. Relative errors and computation time of the finite element method.

| Time step | Number of elements | $\delta\boldsymbol{j}$, % | $\delta\boldsymbol{e}$, % | $\delta\boldsymbol{e}^*$, % | CPU time, min. |
|---|---|---|---|---|---|
| 1e-3 | 1055 | 3.9 | 4.6 | 9.9 | 1.4 |
| 5e-4 | 4418 | 1.8 | 2.1 | 6.6 | 27 |
| 2.5e-4 | 8937 | 1.1 | 1.3 | 4.2 | 300 |

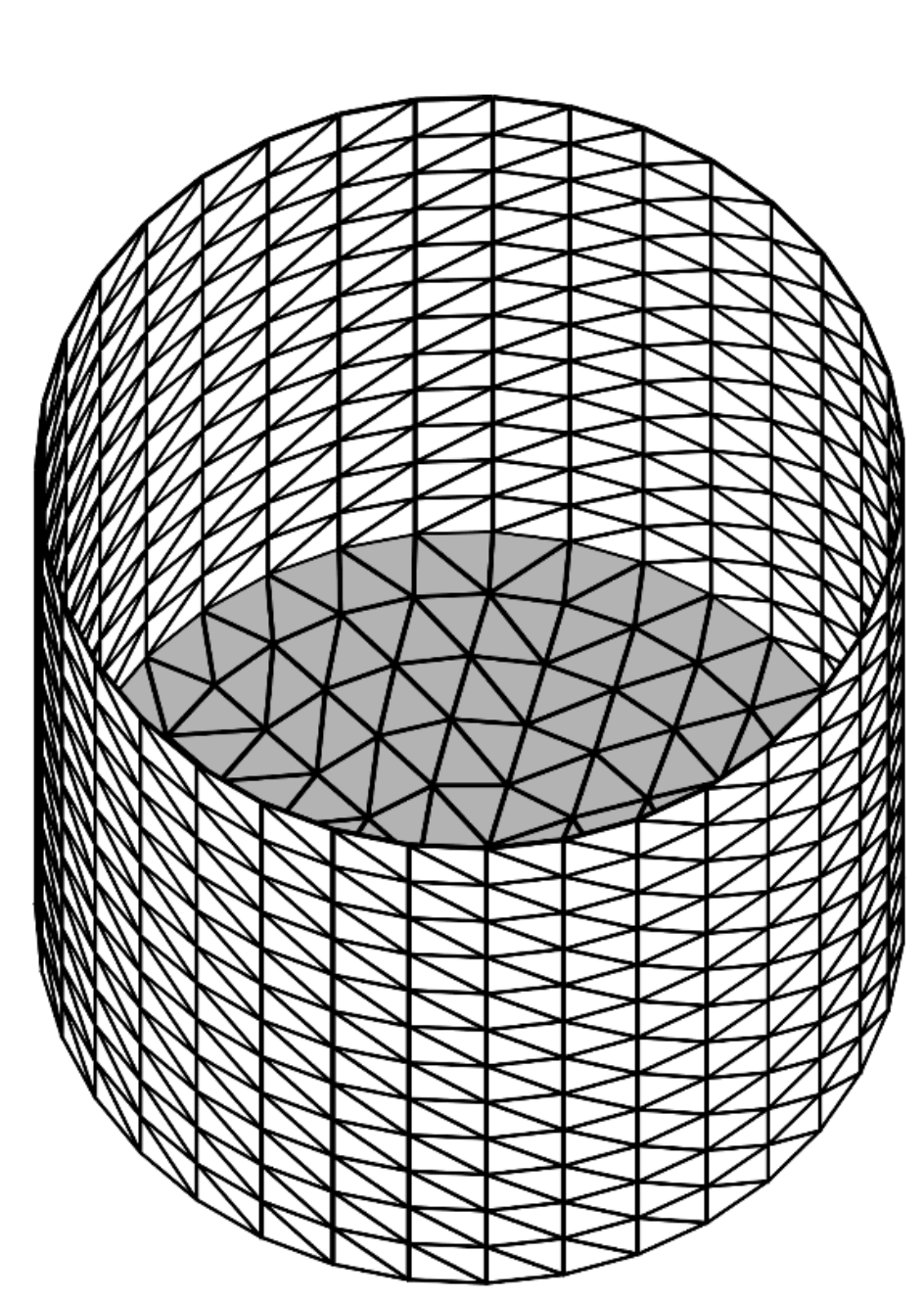

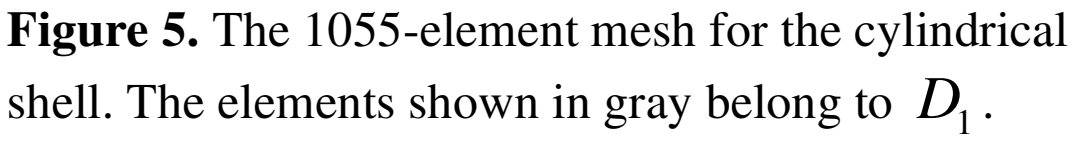

**Figure 5.** The 1055-element mesh for the cylindrical shell. The elements shown in gray belong to $D_1$.

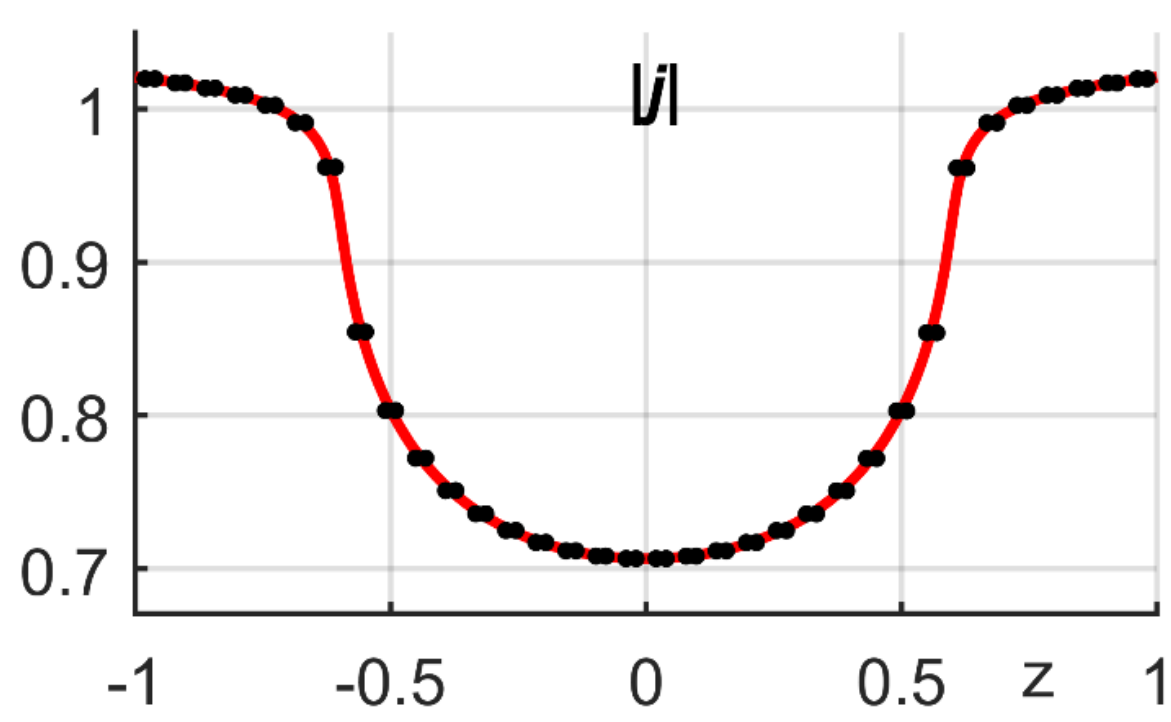

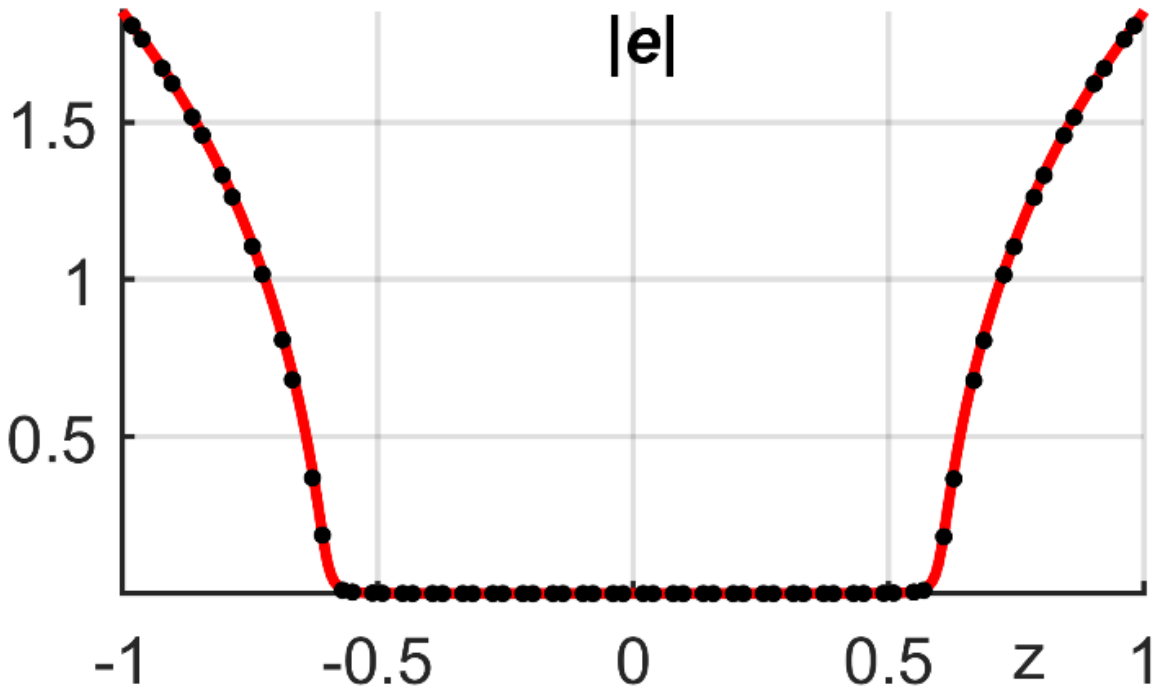

**Figure 6.** Simulation results for the cylindrical surface. Spectral solution (red line) and the finite element solution (black dots) for the mesh with 8937 elements; $n = 30$, $h_z^e = 0.6$.

The examples presented confirm the validity of the developed mixed T-E method for non-flat-film magnetization problems. Our approach provides for a better electric field calculation than the usual direct methods $(\delta e < \delta e^*)$. As the nonlinearity of the current-voltage relation increases, the difference becomes more significant. To demonstrate this, let us consider the power current-voltage relation with $n = 90$, which is a good approximation of the multivalued Bean model relation. Even with this power law, both our mixed and spectral methods efficiently solve, e.g., the hemisphere problem from our first example. Considering the crude mesh now employed (Fig. 1), the achieved accuracy of the mixed method for both variables, $\boldsymbol{j}$ and $\boldsymbol{e}$, is good (Table 4). Our mixed scheme enables determining the electric field for the Bean model as the $n \to \infty$ limit of the power-law model solution. However, due to the high nonlinearity of the current-voltage relation, the direct calculation of the electric field from the sheet current density is unreliable (Fig. 7).

**Table 4.** Hemisphere. Numerical simulation with $n = 90$.

| Time step | Number of elements | $\delta \boldsymbol{j}$, % | $\delta \boldsymbol{e}$, % | $\delta \boldsymbol{e}^*$, % | CPU time, min. |
|---|---|---|---|---|---|
| 1e-3 | 1299 | 2.0 | 4.0 | 21 | 1.6 |

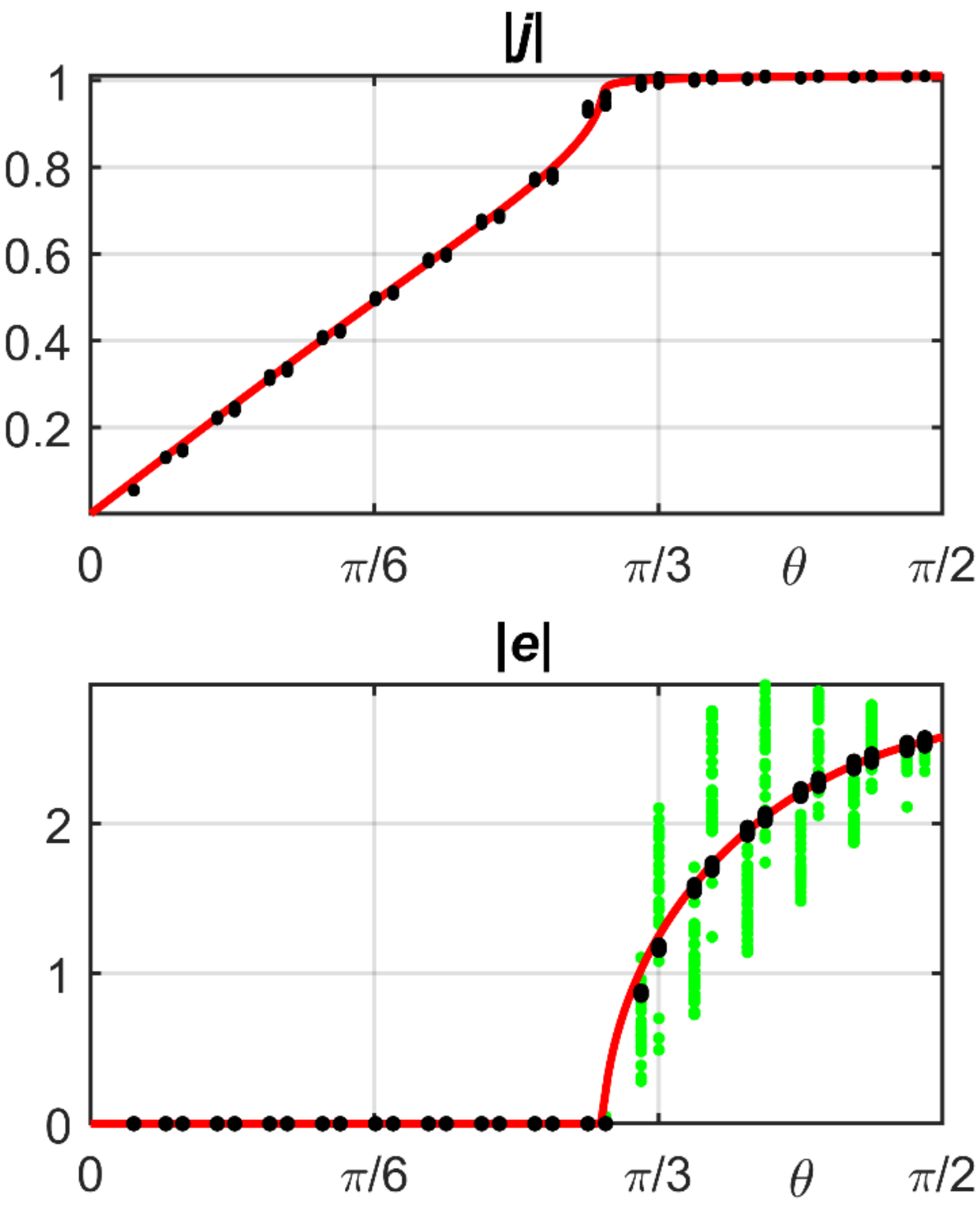

Figure 7. Magnetization of a hemisphere. As in Fig. 1, but for $n$=90 and the 1299-element mesh. Spectral solution (red lines) and the mixed method solution (black dots). The green dots in the bottom plot show the electric field calculated directly.

## 5. Cylindrical HTS magnetic dynamo pump

Magnetic flux pumps are prospective devices capable of inducing strong DC currents in superconducting coils and magnets in a contactless manner, thereby avoiding losses in non-superconducting leads and reducing the cryogenic load [20, 21]. In the dynamo-type pumps, a permanent magnet (PM) moving in the vicinity of an HTS stator creates in it a running magnetic flux wave. Due to the nonlinear resistivity of superconductors, the electric field generated in the stator produces a nonzero time-averaged DC voltage across the stator terminals [22]. Numerical simulations of dynamo pumps, both based on simplified models, as in [23, 24], and upon more realistic ones in [25-28], concerned the pumps with the stator in the form of a coated conductor tape, periodically crossed transversely by one or several rotor-mounted permanent magnets [29, 30].

The cylindrical dynamo pumps in [5, 31, 32] are different: they contain a stator in the form of a thin-walled superconducting cylindrical shell and a PM that rotates with the rotor inside the cylinder, always remaining close to the stator (see Fig. 8).

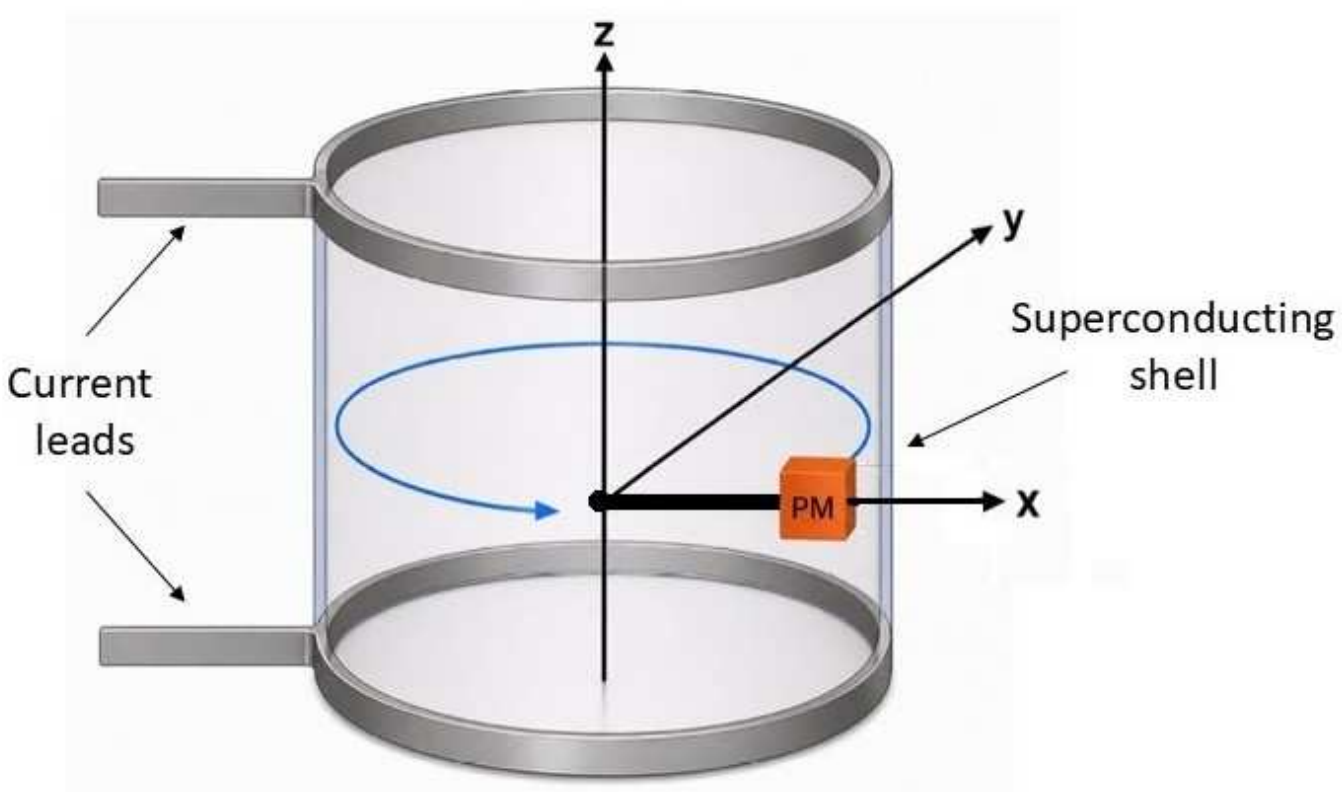

**Figure 8.** A scheme of a cylindrical magnetic dynamo pump.

Using the proposed mixed T-E method, we computed the open-circuit voltage of a cylindrical dynamo. In this example, the dynamo geometry and most parameters (see Table 5) are as in [5]. The PM is uniformly magnetized; its North pole is oriented towards the shell.

We assumed that the field-dependent sheet critical current density is $j_c(\boldsymbol{h}) = j_{c0} / \left(1 + h_0^{-1}\sqrt{|h_\perp|^2 + k_0 |\boldsymbol{h}_\parallel|^2}\right)$, where $j_{c0} = 21.7$ kA/m, $h_0 = 5 j_{c0}$, $k_0 = 0.5$, $h_\perp$ and $\boldsymbol{h}_\parallel$ denote, respectively, the normal and tangential components of the magnetic field on $S$. This field is the sum of the current induced field (15) and the external field produced by the rotating PM; the latter was calculated on each time step analytically (see [33], par. 4.2.1).

In the close-to-PM moving area of the stator, the ratio $|\boldsymbol{j}| / j_c(\boldsymbol{h})$ is high, so the resistivity of the superconducting layer increases, and the substrate current can become significant. Hence, we took into account the finite sheet resistivity of the substrate, $\rho_m$.

**Table 5.** Dynamo model parameters.

| | |
|---|---|
| Cylinder radius | 38 mm |
| Cylinder height | 46 mm |
| Gap between the PM and the shell | 3.7 mm |
| PM dimensions | 10 mm x 10 mm x 10mm |
| PM magnetization | 1.32 T |
| $n$ | 20 |
| $\rho_m$ | $7.58 \cdot 10^{-5}\ \Omega$ |
| Rotor rotation frequency | 25 Hz |

Let us consider two cases (Fig. 9): the full cylindrical stator and the one with a narrow cut (the latter can be easier to realize in practice). If there is no cut, the shell is not simply connected. However, due to the symmetry, the total transport current in the azimuthal direction remains zero, and we can use the mesh as in Fig. 9 (left), setting $T = 0$ on both (upper and lower) parts of $\partial S$. This approach was confirmed by numerical simulations: the same solutions were obtained for the simply connected cylindrical surface (as in Fig. 5) with a weakly conducting bottom.

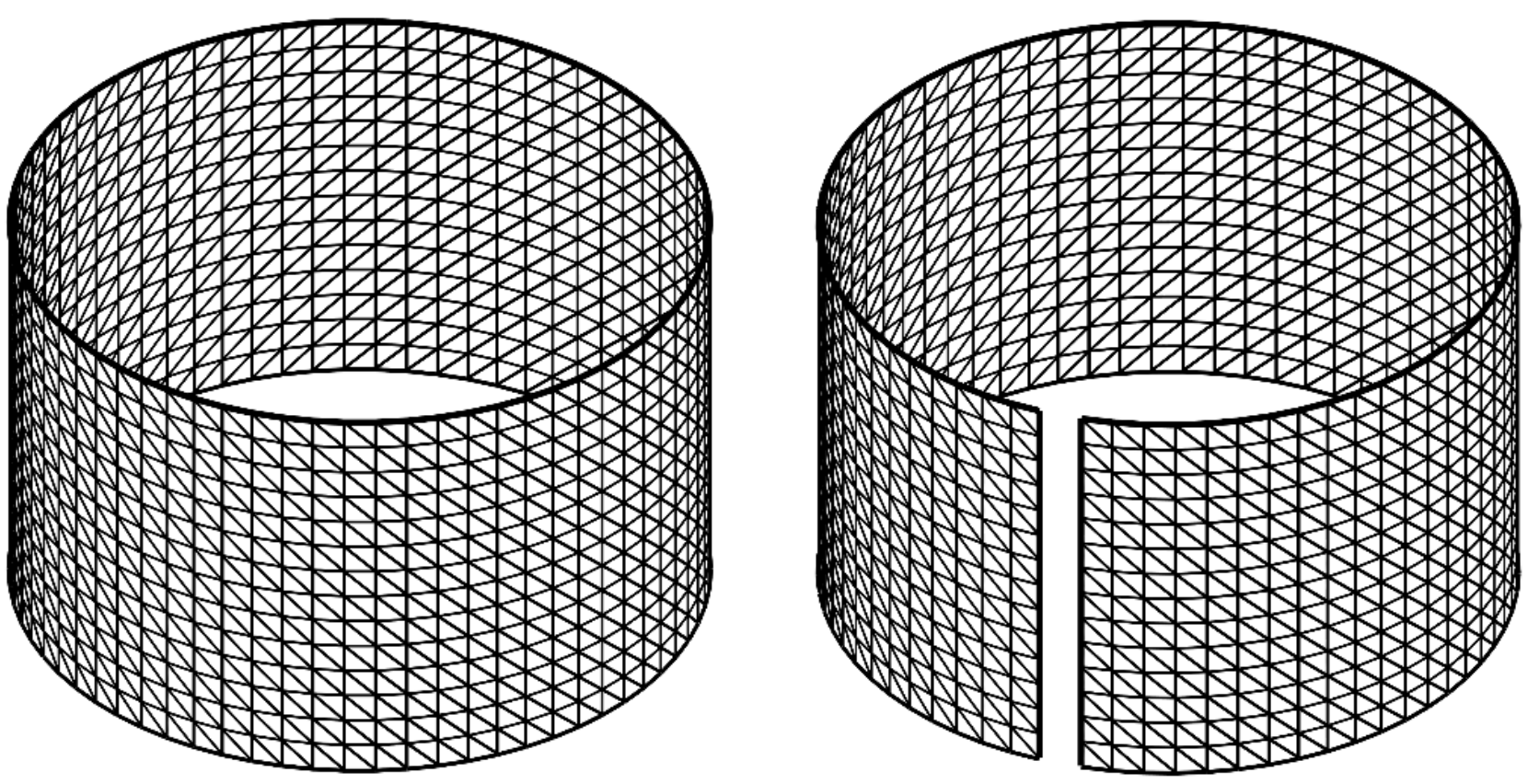

**Figure 9.** Meshes for a cylindrical stator (left) and a stator with a cut (right)

The open-circuit voltage was computed as

$$V(t) = \frac{1}{L}\int_S e_z \mathrm{d}S \,,$$

where $L$ is the length of the upper and lower parts of the shell boundary. Let $R^{cyl}$ denotes the cylinder radius. Then $L = 2\pi R^{cyl}$ for the stator without cuts; if the cut occupies 2% of the cylindrical surface, as in our simulation, $L = 1.96\pi R^{cyl}$. The simulations were started from the virgin initial state.

To evaluate the convergence and accuracy of our scheme for the stator with no cuts, we computed the average open-circuit voltage during the second rotor revolution, $\langle V \rangle$, using different finite element meshes (Table 6); the time step corresponded to the rotor rotation by 3°. The distribution of the sheet current density after the second cycle is shown in Fig. 10, left.

**Table 6.** Stator without cuts. Computation of the open-circuit voltage.

| Number of elements | $\langle V \rangle$, mV | CPU time/cycle, min | $\delta\langle V \rangle$, % |
|---|---|---|---|
| 1960 | 0.5067 | 9 | 3.1 |
| 4160 | 0.4978 | 70 | 1.3 |
| 8642 | 0.4914 | 417 | -- |

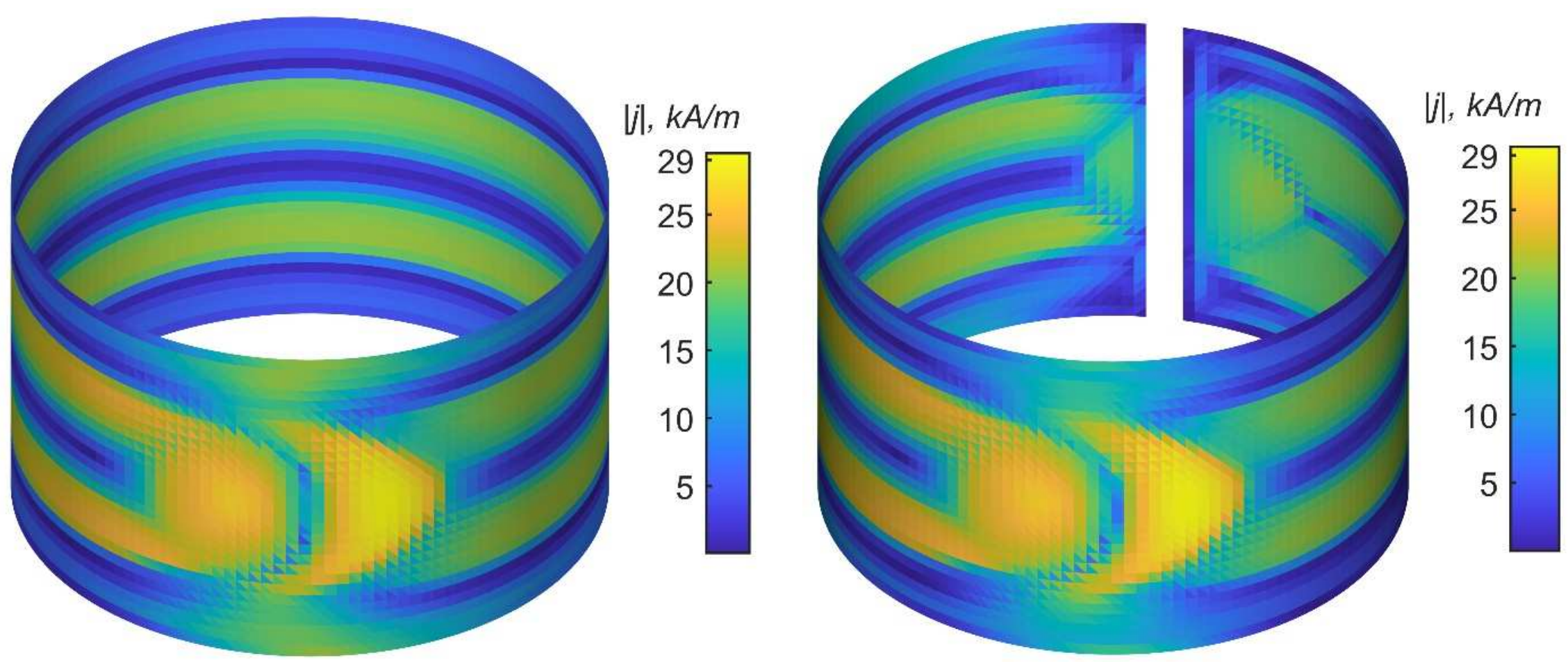

**Figure 10.** Sheet current density after two rotor revolutions: fully cylindrical stator (left), stator with a cut (right).

To clarify the origin of a nonzero DC voltage, let us examine the solution near the PM. Although the positive and negative parts of $j_z$ are, approximately, of the same magnitude, the ratio $j_z / j_c(h)$ is asymmetric (Fig. 11, left). The asymmetry is further magnified by the nonlinearity of the current-voltage relation, and the positive $e_z$ values strongly prevail (Fig. 11, right).

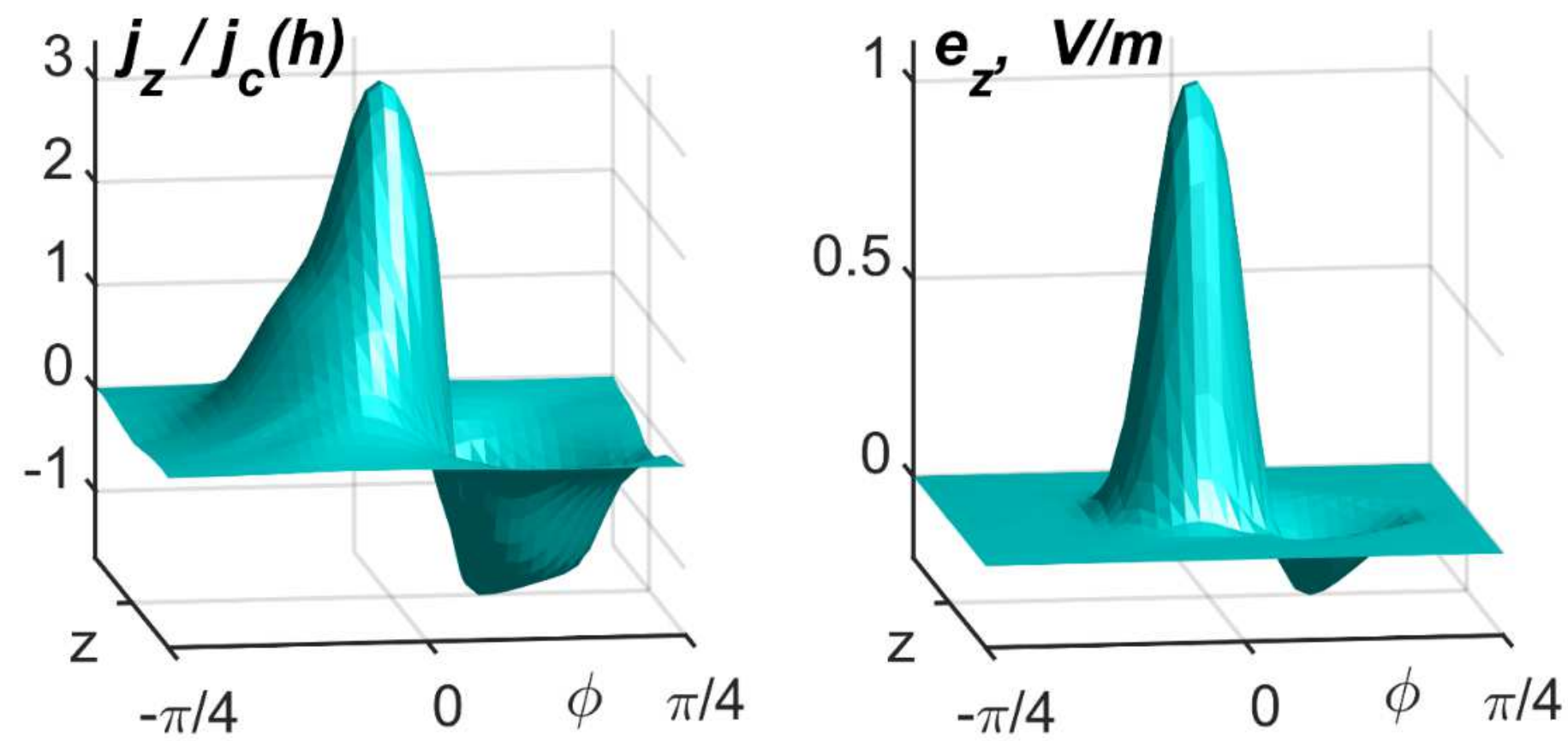

**Figure 11.** Solution near the PM: the positive $e_z$ values prevail. Here $\phi$ is the azimuthal angle of a stator point relative to the PM position.

The stator with a cut produces almost the same average open-circuit voltage during the second rotor revolution (Table 7). The voltage oscillates as the PM passes the cut (Fig. 12). Away from the cut, the sheet current density distribution (Fig. 10, right) is similar to that in a stator without cuts (Fig. 10, left).

**Table 7.** Stator with a cut. Computation of the open-circuit voltage.

| Number of elements | $\langle V \rangle$, mV | CPU time/cycle, min | $\delta\langle V \rangle$, % |
|---|---|---|---|
| 1904 | 0.4941 | 11 | 3.0 |
| 4080 | 0.4856 | 62 | 1.3 |
| 8468 | 0.4795 | 338 | -- |

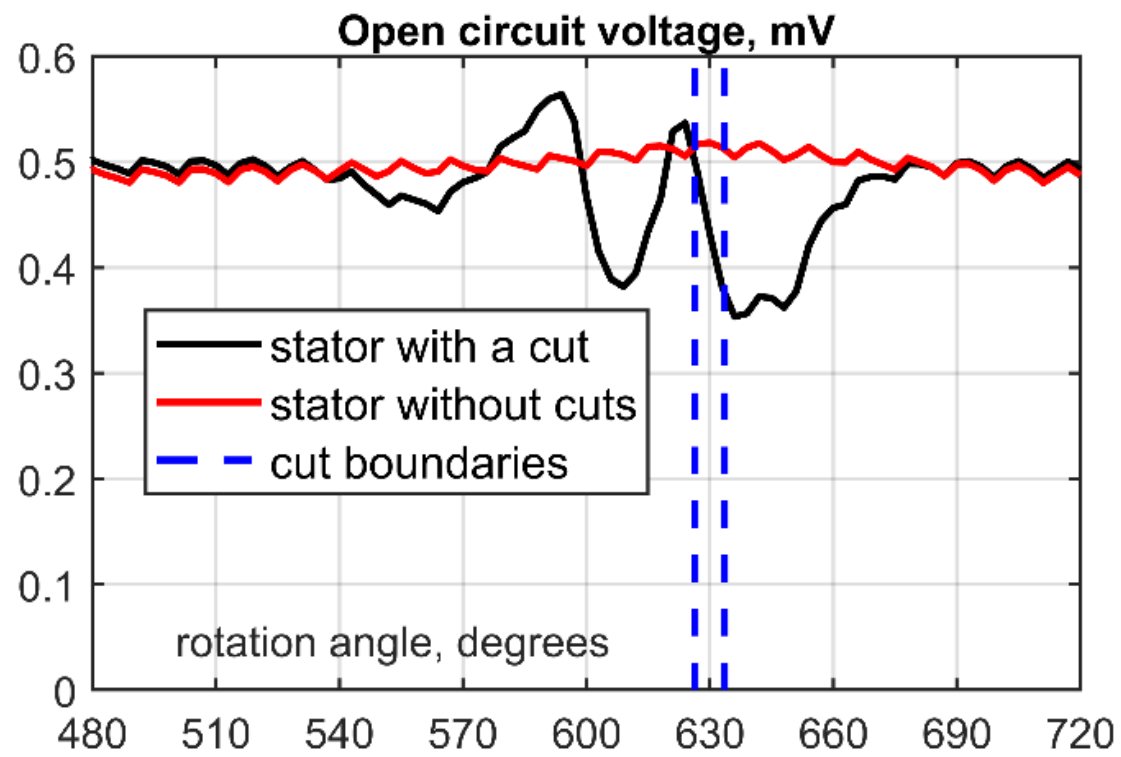

**Figure 12.** Open-circuit voltage oscillates as the PM passes near the cut during the second rotor turn.

For the lower frequency, 9 Hz, the calculated voltage is about 0.2 mV, which is in good agreement with the experiment [5] at a temperature of about 75 K. The value of 0.5 mV obtained at 25 Hz is lower than the experimental value. This difference is probably explained by the heating of the superconductor at a higher frequency.

## 6. Discussion

The integral T-E formulation-based numerical method [15] has been extended to modeling the magnetization of thin non-flat HTS shells. An important special feature of this mixed method is its ability to accurately determine not only the sheet current density but also the electric field.

Unlike most other numerical methods for such problems, our method requires no meshing of the surrounding space, thereby reducing the number of unknowns. It also avoids the complications associated with the film high aspect ratio encountered in the H-formulation-based methods. The finite element approximation we employ ensures the absence of spurious oscillations and uses a simple combination of the vectorial piecewise constant and nonconforming linear elements.

Our approach to solving magnetization problems for multiply connected shells may not be the most efficient or universal. It is, however, applicable to a wide class of shells and is much simpler than the alternative approaches based on advanced algebraic topology tools (see, e.g., [34]).

As with other boundary integral equation formulations, the discretized problems to be solved involve dense matrices. Their elements are double surface integrals, some of which are singular. Well-established analytical and semi-analytical methods exist for computing these integrals (Appendix). Because of the dense matrices, the necessary computer time and memory are badly scaled with the number of unknowns. The methods that resolve this difficulty and allow the use of fine finite element meshes are based on multipole expansions or on hierarchical matrices. These techniques, however, introduce a nontrivial setup cost and implementation overhead. Generally, for problems with fewer than $10^4$ unknowns, the direct approach is preferable. In our examples, the estimated accuracy of numerical solutions was 1-2% for the meshes with a few thousand elements; such accuracy is sufficient for most practical applications.

### Acknowledgment

The surface finite element meshes were generated using custom MATLAB programs developed with assistance from the AI Claude (Anthropic); ChatGPT was used to create Fig. 8.

### Appendix.

For coinciding triangles $\kappa$ and $\kappa'$, the double surface integrals $K_{\kappa,\kappa'}$ were found analytically [35]. For touching and close triangles $\kappa$ and $\kappa'$, the inner integral (over $\kappa'$) of the double surface integral $K_{\kappa,\kappa'}$ was computed analytically (following [36] and using the MATLAB program [37]) at the knots in $\kappa$ of the Gauss quadrature for triangles of the order 30 (see [38, 39]); then the non-singular integral over $\kappa$ was computed numerically using this quadrature. Fully numerical integration using the 5th-order Gauss quadrature for triangles [38, 39] was used for the double surface integrals over distantly separated triangles.

The integrals $\boldsymbol{u}_{\kappa,\kappa'}$ in (16) were computed analytically using the MATLAB implementation [37] of the complicated formulas [36].